\documentclass[10pt]{article}
\usepackage[dvips]{graphicx}
\usepackage{amsmath}
\usepackage{amsthm}
\usepackage{amssymb}
\usepackage{makeidx}

\title{An $R \mid \mid  C_{max} $ Quantum Scheduling Algorithm}

\author{Feng Lu and Dan C. Marinescu\\
 School of Computer Science \\
 University of Central Florida,  P. O. Box 162362\\
 Orlando, FL 32816-2362\\
 Email: { (lvfeng, dcm)@cs.ucf.edu}}

\begin{document}
\maketitle
\date
\begin{abstract}

Grover's search algorithm can be applied to a wide range of
problems; even problems not generally regarded as searching
problems, can be reformulated to take advantage of quantum
parallelism and entanglement, and lead to algorithms which show a
square root speedup over their classical counterparts.

In this paper, we discuss a systematic way to formulate such
problems and give as an example a quantum scheduling algorithm for
an $R||C_{max}$ problem. $R||C_{max}$ is representative for a
class of scheduling problems whose goal is to find a schedule with
the shortest completion time in an unrelated parallel machine
environment.

Given a deadline, or a range of deadlines, the algorithm presented
in this paper allows us to determine if a solution to an
$R||C_{max}$ problem with $N$ jobs and $M$ machines exists, and if
so, it provides the schedule. The time complexity of the quantum
scheduling algorithm is $\mathcal{O}(\sqrt{M^N})$ while the
complexity of its classical counterpart is $\mathcal{O}(M^N)$.

\end{abstract}

\medskip
\section{Introduction}
\medskip

In recent years we have witnessed significant theoretical and some
encouraging experimental results in the area of  quantum
computing. In $1994$, Peter Shor found a polynomial time algorithm
for the factorization of $n$-bit numbers on quantum computers
\cite{Shor94}. His discovery generated a wave of enthusiasm for
quantum computing, for two major reasons: the intrinsic
intellectual beauty of the algorithm and the fact that efficient
integer factorization is a very important practical problem. The
security of widely used cryptographic protocols is based upon the
conjectured difficulty of the factorization of large integers.

Shor's algorithm reduces the factorization problem to the problem
of finding the period of a function, but uses quantum parallelism
to find a superposition of all values of the function in one step.
Then the algorithm calculates the Quantum Fourier Transform of the
function, which sets the amplitudes into multiples of the
fundamental frequency, the reciprocal of the period. To factor an
integer, Shor's algorithm measures the period of the
function\footnote{A powerful version of the technique used by Shor
is the {\it phase-estimation algorithm} of Kitaev
\cite{Kitaev95}}.

In $1996$, Grover described a quantum algorithm for searching an
unsorted database containing $N$ items in a time of order
$\sqrt{N} $ while on a classical computer the search requires a
time of order $N$ \cite{Grover96}. The critical aspect of a
quantum algorithm is to create a superposition of all possible
states and amplify the ``solution''. The speedup of Grover's
algorithm is achieved by exploiting both quantum parallelism and
the fact that in quantum theory a probability is the square of an
amplitude. Bennett and his co-workers \cite{bennett97} and Zalka
\cite{Zalka99} showed that Grover's algorithm is optimal. No
classical or quantum algorithm can solve this problem faster than
time of order $\sqrt{N} $.

Grover's search can be applied to a large number of unstructured
problems and lead to a square root speedup over the corresponding
classical algorithms. For well structured problems classical
algorithms that produce an approximate solution and perform faster
exit.

Recently, Furrow discussed applications based upon Grover-type
algorithms \cite{Furrow06}. He considered three classes of
applications: a) Graph algorithms, e.g. Breadth-First Search
(BFS), Depth-First Search (DFS), bipartite matching. b)
Computational Geometry algorithms, e.g. Maximum points on a line.
c) Dynamic programming algorithms, such as coin changer. The
author reports the quantum versus classical complexity for BFS and
DFS, $\mathcal{O}(\sqrt{VE \lg V})$, versus $\mathcal{O}(E)$, with
$V$ the number of vertices and $E$ the number of edges of the
graph; for bipartite matching, $\mathcal{O}(V \sqrt{(E+V) \lg V
})$, versus $\mathcal{O}((E+V) \sqrt{V}))$; for maximum points
that lie on a line in $\mathbb{R}^{2}$, out of N points,
$\mathcal{O}(N^{3/2}\lg N)$ versus $\mathcal{O}(N^{2}\lg N)$, and
so on.

Most of the problems discussed in \cite{Furrow06} are
intrinsically search problems and the idea of applying Grover's
search comes naturally to mind. There is an even larger class of
problems which, at the first sight, do not seem directly related
to Grover's search. Applications such as scheduling and resource
allocation are not naturally search problems; nevertheless they
share some common properties, can be reformulated to take
advantage of quantum parallelism and entanglement, and lead to
algorithms which show polynomial speedups over their classical
counterparts.

A scheduling problem is characterized by a tuple $(\alpha \mid
\beta \mid \gamma)$ where $\alpha$ denotes the machine
environment, $\beta$ summarizes the set of constraints, and
$\gamma$ denotes the optimality criterion. The makespan of a
schedule, $C_{max}$ is the maximum completion time of any job in
the schedule. For example, $P||C_{max}$ and $R||C_{max}$ require
the shortest makespan and apply to identical machine environment
and, respectively, a non-homogeneous one.

When we turn our attention to problems when a deadline is imposed,
or when we wish to find a schedule with a given range of possible
average completion time  we discover that  a full range of
scheduling problems have a quantum counterpart which can take
advantage of Grover's search.

We illustrate these ideas with an example: given a deadline, or a
range of deadlines, the algorithm presented in this paper allows
us to determine if a solution to an $R||C_{max}$ problem with $N$
jobs and $M$ machines exists, and if so, it provides the schedule.
The time complexity of the quantum scheduling algorithm is
$\mathcal{O}(\sqrt{M^N})$ while the complexity of its classical
counterpart is $\mathcal{O}(M^N)$.

Real-time systems are subject to deadlines and Quality of Service
(QoS)  constraints imposed to many systems require a given range
of average completion times. Thus, the classes of scheduling
algorithms we discuss in this paper are of significant practical
importance. Such algorithms have a quantum counterpart that enjoy
a square root speedup.

\medskip
\section{Scheduling Algorithms}
\label{SchedulingAlgorithms}
\medskip

Scheduling is the problem of assigning tasks to a set of resources
subject to a set of constraints, over time, in an ``optimal''
manner.

We are given a set of $N=2^{n}$ jobs, $\mathcal{J} =\{
\mathcal{J}_{1}, \mathcal{J}_{2}, \ldots \mathcal{J}_{N} \} $ and
a set of $M=2^{m}$ machines, $\mathcal{M} =\{ \mathcal{M}_{1},
\mathcal{M}_{2}, \ldots \mathcal{M}_{M} \} $. A {\it schedule}
$\mathcal{S}$ for the sets $(\mathcal{J}, \mathcal{M} )$ specifies
which $T_{ij}$ units of time machine $\mathcal{M}_{j}$ uses to
process job $\mathcal{J}_{i}$. We call $C_{i}^{\mathcal{S}}$ the
{\it completion time} of job $\mathcal{J}_{i}$ under the schedule
$\mathcal{S}$. The {\it makespan} of the schedule $\mathcal{S}$ is
the maximum completion time of any job in schedule $\mathcal{S}$:

$$
C_{max}^{\mathcal{S}} = \max_{i} C_{i}^{\mathcal{S}}.
$$

We are often interested in scheduling problems involving {\it
multiple} machines.  We distinguish three cases in the {\it
parallel machine environment}:

\begin{enumerate}
\item
{\it Identical parallel environment.} All the machines are
identical and  job $\mathcal{J}_{i}$ requires $T_{i}$ units of
time on any machine,

\item {\it Uniformly related parallel environment.} Each machine
$\mathcal{M}_{j}$ has a speed $s_{j}> 0$ and job $\mathcal{J}_{i}$
if processed entirely on machine $\mathcal{M}_{j}$ would take $
T_{i}/s_{j}$ units of time, and

\item
{\it Unrelated parallel machine environment.} Different machines
have different capabilities and the speed of machine
$\mathcal{M}_{j}$ on job $\mathcal{J}_{i}$ is $s_{ij}$. Then the
processing time of job $\mathcal{J}_{i}$ on machine
$\mathcal{M}_{j}$ is $T_{ij} = T_{i}/s_{ij}$, $0 \le T_{ij} \le Q$
and $Q=2^{q}$.

\end{enumerate}

\noindent These machine environment are denoted by $P, Q,$ and
$R$, respectively.

\medskip

Examples of scheduling constraints include deadlines (e.g., job
$i$ must be completed by time $t$), resource capacities (e.g.,
there are only $5$ machines), precedence constraints on the order
of tasks (e.g., one must be done before another), and priorities
on tasks (e.g., finish job $j$ as soon as possible while meeting
the other deadlines). A {\it priority rule} assigns to job
$\mathcal{J}_{i}$ a priority $\pi_{i}$. A {\it busy schedule}
defines the situation where when one machine becomes available it
starts processing the job with the highest priority.

\medskip

Each scheduling problem could have problem specific optimization
criteria. The one which requires the minimization of the makespan
is referred to in scheduling theory as $C_{max}$. Other
optimization criteria can be considered. For example, we may wish
to optimize the average completion time of all jobs:

$$
{1 \over N} \sum_{i=1}^{N} C_{i}^{\mathcal{S}}
$$
an optimization criterion denoted as $\sum_{i=1}^{N}
C_{i}^{\mathcal{S}}$.

Many scheduling problems are $\mathcal{N}\mathcal{P}-$ hard on
classical computers \cite{Brucker06} and have proved to be very
difficult even for a relatively small instance. For example, a
10-job-10-machine job-shop scheduling problem posed in $1963$
remained unsolved until $1989$ \cite{Carlier89}. Most classical
scheduling algorithms gain speedup only on some special cases with
relatively small instance or they try to find good schedules
instead of the optimal one. Furrow\cite{Furrow06} also gave a
special case of the $P||C_{max}$ problem that can be solved by
dynamic programming, which requires that the number of different
jobs processing times be bounded by a constant. It turns out that
there are polynomial approximations for some $P||C_{max}$
problems.

We consider an $R||C_{max}$ scheduling problem in which all
machines are unrelated. All jobs are available at the beginning
time and there are no precedence constraints. As no preemption is
allowed, once job $\mathcal{J}_{i}$ started processing on machine
$\mathcal{M}_{j}$ it must complete its execution before another
job, $\mathcal{J}_{k}$ can be processed on $\mathcal{M}_{j}$.

Given a set of $N=2^n$ jobs and $M=2^m$ machines we can construct
$2^{m2^n}$ different schedules. This $R||C_{max}$ scheduling
problem is $\mathcal{N}\mathcal{P}-$ hard and is difficult for
classical algorithms even with small instance. Up to now, most
classical algorithms use linear programming based rounding
techniques(LP-rounding) to search an approximate schedule instead
of the optimal one \cite{Grigoriev05,Schulz02}. If the number of
machines $m$ is part of the input, the best approximation
algorithm to date is a 2-approximation by Lenstra, Shmoys and
Tardos \cite{Lenstra90} which can find a schedule with
$C_{max}<2*C^{opt}_{max}$. Moreover, the problem cannot be
approximated within a factor strictly smaller than 3/2, unless
P=NP \cite{Lenstra90}. No proper classical algorithm addresses a
general case of searching the optimal schedule except the
exhausting search which has time complexity $\mathcal{O}(M^N)$ on
a classical computer. The paper suggests a reformulation of such
problems to take advantage of Grover-type search and gain square
root speedup on searching the optimal schedules over their
classical counterparts.

 \medskip
 \section{Information Encoding}
 \label{InformationEncoding}
 \medskip

Consider a set of $N=2^n$ jobs running on $M=2^m$ unrelated
machines. We assume that the jobs could have different processing
times on different machines and that the processing times are
integers in the range $0 \le T_{ij} < Q=2^q, ~~~  1 \le i \le
2^{n}, ~~~1 \le j \le 2^{m}$. We also assume that we have a
quantum system with $r=m+q$ qubits for each job.

Given job $\mathcal{J}_{i}$ running on machine $\mathcal{M}_{j}$,
we encode the {\it job-machine} information as a vector $\mid
e_i^j \rangle$ obtained as the tensor product of the machine
index, $j$, and the processing time, $T_{ij}$:

$$
\mid e_i^j \rangle = \mid j \rangle \otimes \mid T_{ij} \rangle.
$$
Then we define {\it job state vectors} for job $i$, $\mid J_{i}
\rangle$ as any superposition of its job-machine vectors. First,
we consider an equal superposition of job-machine vectors as:

$$
\mid J_{i} \rangle = { 1 \over \ {2^{m/2}} }  \sum_{j=1}^{2^m}
\mid e_{i}^{j} \rangle.
$$
A {\it schedule} is a tensor product of job-machine vectors:

$$
\mid \mathcal{S}_{k}\rangle = \otimes_{i=1}^{2^n} \mid e_{i}^{j_i}
\rangle,~~~ 1 \le j_i \le 2^m,
$$
which includes one job-machine vector for each job. A specific
machine may be present in multiple job-machine vectors, when the
schedule requires multiple jobs to be executed on the same
machine, or may not appear in any job machine vectors of the
schedule $\mid \mathcal{S}_{k}\rangle$ if none of the jobs is
scheduled on that machine.

The equal superposition of all schedules is:

$$
\mid \mathcal{S} \rangle = {1 \over \sqrt{\sigma}}
\sum_{k=1}^{\sigma} \mid \mathcal{S}_{k} \rangle ,~~~\sigma= 2^{m
2^{n}}
$$

Let $\Omega$ be an operator which given a schedule constructs the
running time on each machine for that schedule. When applied to
the equal superposition of all schedules, it produces a
superposition of the running time $\mid \mathcal{T}\rangle$ on
each machine for all schedules:

$$
\mid \overbrace{\mathcal{S}\mathcal{T}} \rangle = \Omega (\mid
\mathcal{S} \rangle\mid 0\rangle).
$$
where, $\mid \overbrace{\mathcal{S}\mathcal{T}}\rangle$ denotes
the entangled state of $\mathcal{S}$ and $\mathcal{T}$, while the
tensor product of $\mathcal{S}$ and $\mathcal{T}$  is  $\mid
\mathcal{S} \mathcal{T}\rangle$.

Let $\Delta$ be an operator which computes a superposition of the
makespan of all schedules:

$$
\mid \overbrace{\mathcal{S}\mathcal{T}C_{max}} \rangle  = \Delta
(\mid \overbrace{\mathcal{S}\mathcal{T}}\rangle \mid 0\rangle)=
\Delta ~ \Omega  (\mid \mathcal{S}\rangle\mid 0\rangle\mid
0\rangle)  .
$$

\begin{figure}[h]
\begin{center}
\includegraphics[width=8cm]{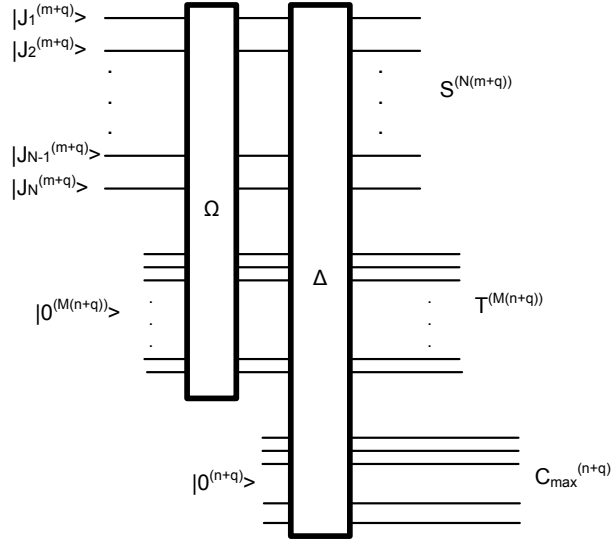}
\end{center}
\caption{Outline of our algorithm to prepare the makespan vector.
First we prepare job vectors in an equal superposition of all
possible schedules $\mid \mathcal{S}^{(N(m+q))} \rangle$. Using a
quantum circuit $\Omega$ with inputs $\mid \mathcal{S}^{(N(m+q))}
\rangle$ and a register of $M(n+q)$ qubits in state $\mid 0
\rangle$, we construct the superposition of the running time on
each machine for every schedule. Then we construct the makespan of
each schedule using the operation $\Delta$. Superscripts indicate
the number of qubits for vectors. Note that the number of jobs is
$N=2^{n}$, the number of machines is $M=2^{m}$, and the maximum
execution time of a job on any machine is $Q=2^{q}$. }
 \label{makespan}
\end{figure}

Figure \ref{makespan} outlines our procedure to produce an equal
superposition of the makespans of all schedules. We now turn our
attention to the quantum circuits to carry out the transformations
discussed in this section and to the question how to obtain from
the superposition of all makespans the optimal one.

\medskip

\begin{figure}[h]
\begin{center}
\includegraphics[width=7.5cm]{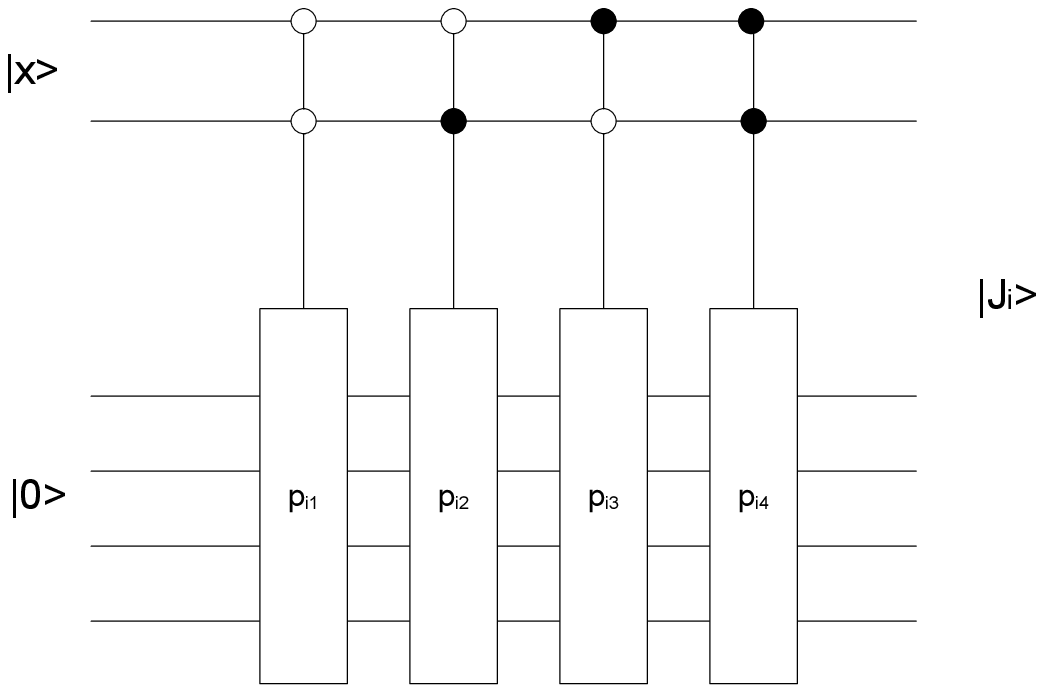}
\end{center}
\caption{A quantum circuit to prepare the job state vectors with
$m=4$. In this case $\mid x \rangle$ is a set of two control
qubits. Each control qubit is set to $ {1 / \sqrt{2}}(\mid 0
\rangle+ \mid 1 \rangle)$. The target qubits will be prepared in a
superposition of all possible job-machine states.} \label{pre}
\end{figure}

First, we need to prepare the job vector $ \mid J_{i} \rangle $ in
an equal superposition state which includes the processing times
of job $\mathcal{J}_{i}$ on all machines, as shown in Figure
\ref{pre}. We use $m$ index qubits to control the job-machine
information encoding. As each index qubit is prepared in state $
{1 /\sqrt{2}}( \mid 0\rangle+\mid 1\rangle)$, the target qubits
will be prepared in superpositions of all possible job-machine
states, $e_{i}^{j},~~~ 1 \le i \le n, ~ 1 \le j \le m$.

\medskip

\noindent {\bf Example:}  Table \ref{executionTime} summarizes the
processing time of $8$ jobs on $4$ machines, where $0 < T_{ij} <
2^{4}=16$.  Thus $n=3$, $m=2$, and $q=4$. The running time of
$\mathcal{J}_{1}$ on machines $\mathcal{M}_{1}, \mathcal{M}_{2},
\mathcal{M}_{3}$ and $ \mathcal{M}_{4}$ are respectively, $1, 3,
7,$ and $15$ units of time.

\begin{table}[h]
\caption{The processing time of $8$ jobs on $4$ machines}
\begin{center}
\begin{tabular} {c||cccc||c||cccc}
 \hline
 Job/Machine & $\mathcal{M}_1$ & $\mathcal{M}_2$ &  $\mathcal{M}_3$  & $\mathcal{M}_4$ &Job/Machine & $\mathcal{M}_1$ & $\mathcal{M}_2$ &  $\mathcal{M}_3$  & $\mathcal{M}_4$\\
 \hline \hline
 $\mathcal{J}_1$ &  1 & 3  & 7 & 15 & $\mathcal{J}_5$ & 15 & 12 & 3 & 10 \\
 $\mathcal{J}_2$ &  2 &  1 & 9 &  3 & $\mathcal{J}_6$ & 10 &  7 & 8 & 14 \\
 $\mathcal{J}_3$ &  6 &  2 & 5 &  8 & $\mathcal{J}_7$ &  5 &  2 & 3 &  9 \\
 $\mathcal{J}_4$ & 11 & 13 & 7 &  4 & $\mathcal{J}_8$ &  1 & 10 &11 &
 13\\
 \hline
\end{tabular}
\end{center}
\label{executionTime}
\end{table}
The four vectors used to encode the processing time of job
$\mathcal{J}_{1}$ on machines $\mathcal{M}_{1}, \mathcal{M}_{2},
\mathcal{M}_{3}$ and $ \mathcal{M}_{4}$ are respectively:

$$
 \mid e_1^1 \rangle = \mid 000001 \rangle, ~~~~\mid e_1^2 \rangle =  \mid 010011
 \rangle, ~~~~
 \mid e_1^3 \rangle = \mid 100111 \rangle, ~~~~\text{and}~~~~\mid e_1^4  \rangle =\mid 111111 \rangle.
 $$
 For $\mathcal{J}_{2}$  the basis vectors are $\mid 000010 \rangle$, $\mid 010001
\rangle$, $\mid 101001 \rangle, \mid 110011 \rangle$, and so on.

\begin{figure}[h]
\begin{center}
\includegraphics[width=7.5cm]{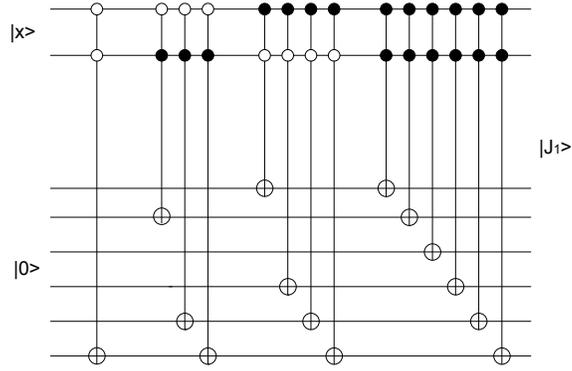}
\end{center}
\caption{A quantum circuit to prepare $\mid J_{1} \rangle$. The
execution times of $J_{1}$ are  1, 3, 7, and 15 units of time
respectively. $\mid x \rangle$ are control qubits which control
the selection of the four job-machine states, $e_{1}^{j},~~ 1 \le
j \le 4$. The first two qubits of $ \mid J_1\rangle$ record the
machine index, and the remaining four qubits  record the time if
$J_1$ is assigned to the corresponding machine.}
 \label{pre1}
\end{figure}

Figure \ref{pre1} shows the circuit used to prepare the job vector
$\mid J_{1} \rangle$ for our example. The job state  $\mid J_{1}
\rangle$ is prepared in an equal superposition of basis states:

$$
\mid J_{1} \rangle= {{1} \over {2}} ( \mid 000001 \rangle + \mid
010011 \rangle + \mid 100111 \rangle + \mid 111111 \rangle).
$$

We can prepare other job vectors in the same way, for example:
$$
\mid J_{2} \rangle= {{1} \over {2}} ( \mid 000010 \rangle + \mid
010001 \rangle + \mid 101001 \rangle + \mid 110011 \rangle).
$$
\medskip

A schedule vector $\mid \mathcal{S} \rangle$ is the tensor product
of all job vectors. As each job vector is prepared as an equal
superposition, the schedule vector is in the equal superposition
of all possible schedules.

We now provide two examples of schedule vectors:

\medskip

\noindent {\bf (i)} Schedule $\mathcal{S}_1$:

$$
 [
 \mathcal{J}_{1} \mapsto \mathcal{M}_{1},~~
 \mathcal{J}_{2} \mapsto \mathcal{M}_{2},~~
 \mathcal{J}_{3} \mapsto \mathcal{M}_{1},~~
 \mathcal{J}_{4} \mapsto \mathcal{M}_{4},~~
 \mathcal{J}_{5} \mapsto \mathcal{M}_{3},~~
 \mathcal{J}_{6} \mapsto \mathcal{M}_{2},~~
 \mathcal{J}_{7} \mapsto \mathcal{M}_{3},~~
 \mathcal{J}_{8} \mapsto \mathcal{M}_{1}]
$$

Schedule $\mathcal{S}_{1}$ corresponds to the following job state
vectors:

\medskip

$\mid J_1 \rangle= \mid \underline{00}0001 \rangle$ \hspace{0.3in}
$\mid J_2 \rangle= \mid \underline{01}0001 \rangle$\hspace{0.3in}
$\mid J_3 \rangle= \mid \underline{00}0110 \rangle$\hspace{0.3in}
$\mid J_4 \rangle= \mid \underline{11}0100 \rangle$

$\mid J_5 \rangle= \mid \underline{10}0011 \rangle$ \hspace{0.3in}
$\mid J_6 \rangle= \mid \underline{01}0111 \rangle$\hspace{0.3in}
$\mid J_7 \rangle= \mid \underline{10}0011 \rangle$\hspace{0.3in}
$\mid J_8 \rangle= \mid \underline{00}0001 \rangle$\medskip

\medskip

\noindent The schedule vector for schedule $\mathcal{S}_{1}$ is:

$\mid \mathcal{S}_{1}\rangle = \mid \underline{00}0001
\rangle\otimes \mid \underline{01}0001 \rangle\otimes \mid
\underline{00}0110 \rangle\otimes \mid \underline{11}0100 \rangle$

\hspace{1.0in}$\otimes \mid \underline{10}0011 \rangle\otimes \mid
\underline{01}0111 \rangle\otimes \mid \underline{10}0011
\rangle\otimes \mid \underline{00}0001 \rangle$

\noindent The completion times on all machines are:\medskip

$
    C^{{A}_{1}}(\mathcal{M}_{1}) =1+6+1=8,
 ~~~C^{{A}_{1}}(\mathcal{M}_{2}) =1+7=8,
 ~~~C^{{A}_{1}}(\mathcal{M}_{3}) =3+3=6,
 ~~~C^{{A}_{1}}(\mathcal{M}_{4}) =4.
$\medskip

The makespan of schedule $\mathcal{S}_{1}$ is equal to the largest
completion time over all machines:

$$
C_{max}^{A_{1}} =8.
$$

\medskip

\noindent {\bf (ii)} Schedule $\mathcal{S}_{2}$:

$$
 [
 \mathcal{J}_{1} \mapsto \mathcal{M}_{2},~~
 \mathcal{J}_{2} \mapsto \mathcal{M}_{1},~~
 \mathcal{J}_{3} \mapsto \mathcal{M}_{3},~~
 \mathcal{J}_{4} \mapsto \mathcal{M}_{4},~~
 \mathcal{J}_{5} \mapsto \mathcal{M}_{3},~~
 \mathcal{J}_{6} \mapsto \mathcal{M}_{2},~~
 \mathcal{J}_{7} \mapsto \mathcal{M}_{1},~~
 \mathcal{J}_{8} \mapsto \mathcal{M}_{2}]
$$

\medskip

\noindent The schedule vector is:\medskip

$\mid \mathcal{S}_{2}\rangle =
 \mid \underline{01}0011 \rangle\otimes
 \mid \underline{00}0010 \rangle\otimes
 \mid \underline{10}0101 \rangle\otimes
 \mid \underline{11}0100 \rangle$

 \hspace{1.0in}
 $\otimes \mid \underline{10}0011 \rangle
  \otimes \mid \underline{01}0111 \rangle
  \otimes \mid \underline{00}0101 \rangle
  \otimes \mid \underline{01}1010 \rangle)$

\bigskip

The schedule vector can also be in a superposition of some basic
states, for example, the schedule vector could be in an equal
superposition of the two schedules, $\mathcal{S}_1$ and
$\mathcal{S}_2$:\medskip

$\mid \mathcal{S}\rangle= 1/\sqrt{2}(\mid
\mathcal{S}_1\rangle+\mid \mathcal{S}_2 \rangle)$

\hspace{0.3in}$=1/\sqrt{2}
 ( \mid \underline{00}0001 \rangle \otimes
   \mid \underline{01}0001 \rangle \otimes
   \mid \underline{00}0110 \rangle \otimes
   \mid \underline{11}0100 \rangle$

\hspace{1.0in}$\otimes
 \mid \underline{10}0011 \rangle \otimes
 \mid \underline{01}0111 \rangle \otimes
 \mid \underline{10}0011 \rangle \otimes
 \mid \underline{01}0001 \rangle$

\hspace{0.3in}$+
 \mid \underline{01}0011 \rangle \otimes
 \mid \underline{00}0010 \rangle \otimes
 \mid \underline{10}0101 \rangle \otimes
 \mid \underline{11}0100 \rangle$

\hspace{1.0in}$\otimes
 \mid \underline{10}0011 \rangle \otimes
 \mid \underline{01}0111 \rangle \otimes
 \mid \underline{00}0101 \rangle \otimes
 \mid \underline{01}1010 \rangle).$

\medskip
\section{The Running Time of Machine $\mathcal{M}_{j}$ under Schedule $\mathcal{S}_{k}$}
\label{RunningTime}
\medskip

Computing the total running time of machine $\mathcal{M}_{j}$
under schedule $\mathcal{S}_{k}$ requires summation of the running
times of all jobs assigned to $\mathcal{M}_{j}$. A quantum adder
similar to a classic adder, e.g. $\mid 5\rangle \hat{+}\mid
6\rangle = \mid 11\rangle$, has:

\begin{itemize}

\item
$n$ inputs $a_1, a_2, \cdots, a_n$, each one is a register of $q$
qubits,

\item
one carry in $c$, and

\item
one output $S=\sum_{i=1}^n a_i+c$, a register of $q+n$ qubits.
\end{itemize}

\medskip

To obtain the total running time of machine $\mathcal{M}_j$ under
schedule $\mathcal{S}_{k}$ we add the execution times, (in our
examples those qubits of a job-machine vector which are not
underlined) for all jobs assigned to $\mathcal{M}_j$ and create a
running time vector for machine $\mathcal{M}_j$, $\mid T_j
\rangle$. The {\it running time vector for  schedule}
$\mathcal{S}_{k}$ is the tensor product of the execution time on
individual machines under schedule $\mathcal{S}_{k}$:

$$
\mid \mathcal{T}^{\mathcal{S}_{k}} \rangle=
 \mid T_1^{\mathcal{S}_{k}} \rangle \otimes
 \mid T_2^{\mathcal{S}_{k}} \rangle \cdots\otimes
 \mid T_M^{\mathcal{S}_{k}} \rangle.
$$

For example, for machine $\mathcal{M}_1$ under schedule
$\mathcal{S}_1$ (see Section \ref{InformationEncoding}):

 $$
 \mid T_1^{\mathcal{S}_1}\rangle:~~~~~
 \mid 0001\rangle \hat{+}  \mid 0000\rangle  \hat{+}\mid 0110\rangle \hat{+}
 \mid 0000\rangle \hat{+} \mid 0000\rangle \hat{+} \mid 0000\rangle \hat{+} \mid 0000\rangle
 \hat{+} \mid 0001\rangle,
 $$
 or

$$
 \mid T_1^{\mathcal{S}_1}\rangle:~~~~~ \mid 0001000\rangle.
$$

\begin{figure}[h]
\begin{center}
\includegraphics[width=10cm]{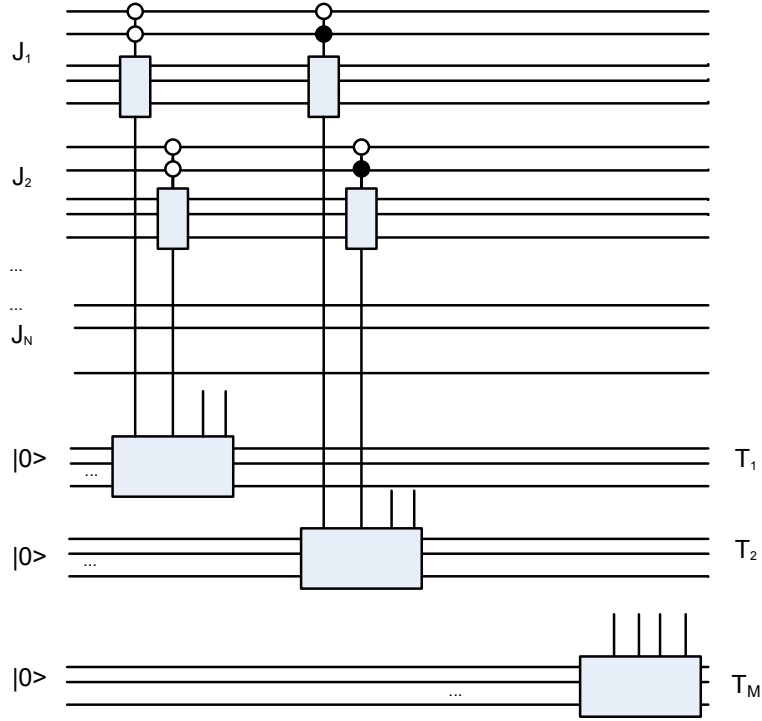}
\end{center}
\caption{A circuit to compute the sum of the execution time of
jobs assigned to each machine. $J_1, J_2 \ldots J_N$ are the job
vectors prepared according to the scheme discussed in Section
\ref{InformationEncoding}, and $T_1, T_2, \ldots T_M$ represent
the execution time on machines
$\mathcal{M}_1,\mathcal{M}_2,...\mathcal{M}_{M}$ respectively.}
 \label{Summation1}
\end{figure}

We want to construct a quantum circuit to sum the execution time
of jobs assigned to each machine $\mathcal{M}_{j}$. We use the
index qubits as control qubits to control the summation of each
index, or each machine; the index qubits are entangled with the
target qubits which give the running time on that machine. As the
input qubits $\mid \mathcal{S}\rangle$ are prepared in the equal
superposition of all possible schedules, this operation will
prepare the running time of machines under all possible schedules.

In Figure \ref{Summation1}, $T_1, T_2, \ldots T_M$ represent the
execution time on machines
$\mathcal{M}_1,\mathcal{M}_2,...\mathcal{M}_{M}$ respectively.
When the index qubits are not ``active'', the respective input for
the summation circuit will be zero.  For example, if the index of
$e_2^2=\mid \underline{01}0001 \rangle$ is $\underline{01}$,
rather than $\underline{00}$, then the index qubits of $e_2^2$ for
$\mid T_1 \rangle$ are not active.\medskip

How to implement arithmetic on a quantum computer has been address
by many authors. Many detailed quantum circuits and discussions
can be found in \cite{Gossett98,Draper04,Meter05}. For the example
discussed in Section \ref{InformationEncoding}, after the
summation operation, the system will be in the state:

\medskip

 $$
 \mid \overbrace{\mathcal{S}\mathcal{T}}\rangle =
 1/\sqrt{2}(\mid \mathcal{S}_1 \mathcal{T}^1 \rangle +
            \mid \mathcal{S}_2 \mathcal{T}^2 \rangle)
 $$
or

 \hspace{0.3in}$\mid \overbrace{\mathcal{S}\mathcal{T}}\rangle =1/\sqrt{2}( \mid \underline{00}0001
\rangle\otimes \mid \underline{01}0001 \rangle\otimes \mid
\underline{00}0110 \rangle\otimes \mid \underline{11}0100
\rangle\otimes \mid \underline{10}0011 \rangle\otimes \mid
\underline{01}0111 \rangle$

\hspace{1.0in}$~\otimes \mid \underline{10}0011 \rangle\otimes
\mid \underline{01}0001 \rangle\otimes\mid
0001000\rangle\otimes\mid 0001000\rangle\otimes\mid
0000110\rangle\otimes\mid 0000100\rangle $

\hspace{1.0in}$~+ \mid \underline{01}0011 \rangle\otimes \mid
\underline{00}0010 \rangle\otimes \mid \underline{10}0101
\rangle\otimes \mid \underline{11}0100 \rangle\otimes \mid
\underline{10}0011 \rangle\otimes \mid \underline{01}0111 \rangle$

\hspace{1.0in}$~\otimes \mid \underline{00}0101 \rangle\otimes
\mid \underline{01}1010 \rangle\otimes\mid
0000111\rangle\otimes\mid 0010100\rangle\otimes\mid
0001000\rangle\otimes\mid 0000100\rangle)$

\medskip
\section{Determination of the Makespan}
\label{Makespan}
\medskip

\begin{figure}[h]
\begin{center}
\includegraphics[width=7cm]{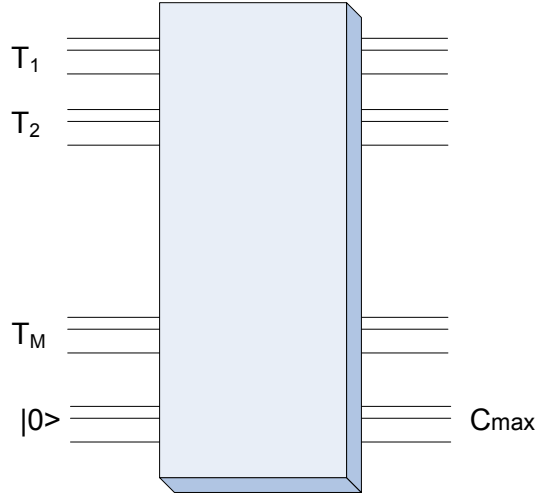}
\end{center}
\caption{A quantum circuit to compute the makespan, the maximum
running time among all machines. The output of this circuit is an
entangled state.} \label{Max}
\end{figure}

Now we have the system in state $\mid
\overbrace{\mathcal{S}\mathcal{T}}\rangle$, of which the last
$M(n+q)$ qubits provide the running time of all $M$ machines under
all schedules. The makespan of a schedule is equal to the maximum
running time among the $M$ machines under that schedule. We want
to construct a {\tt Max} circuit, as shown in Figure \ref{Max}, to
compute the makespan of each schedule. The quantum circuit
computes the maximum over an input set,  e.g., $\text{Max} (\mid
5\rangle, \mid 7\rangle, \mid 4\rangle)=\mid 7\rangle$. The input
to this circuit is a set of $n+q$ qubits. The output $\mid
\text{Max} \rangle$, has also $n+q$ qubits. Implementing such
arithmetic on a quantum computer has been addressed by many
researchers\cite{Gossett98,Draper04,Meter05} and we omit the
detail circuit here.

The output of a quantum circuit as in Figure \ref{Max} is an
entangled state, rather than the tensor product  $\mid
T_1\rangle\otimes \mid T_2\rangle\otimes \cdots \otimes \mid
T_M\rangle\otimes\mid C_{max}\rangle$. Recall that the $\mid
\mathcal{T}\rangle$ was prepared in an equal superposition of
running times of the machines under all possible schedules. Thus,
such a {\tt Max} operation prepares the $C_{max}$ register in an
equal superposition of the makespans of all possible schedules.

\medskip

Up to now, we discussed the implementation of the quantum circuits
presented in Figure \ref{makespan}. During these operations, we
successfully entangle the job vectors with the {\tt Sum} and the
{\tt Makespan} vectors. These vectors are prepared in the equal
superposition of all possible schedules, which can be written as:

$$
\mid \overbrace{\mathcal{S}\mathcal{T}C_{max}} \rangle  =
\sum_1^{2^{m2^n}}\mathcal{S}_i\mathcal{T}_iC_{maxi},
$$
where, $1\leq i \leq2^{m2^n}$ indexes different schedules, and we
ignore the coefficients.

\medskip

For the simple example in Section \ref{InformationEncoding}, the
system will be in the state :

$\mid \overbrace{\mathcal{S}\mathcal{T}C_{max}} \rangle =
1/\sqrt{2}(\mid \mathcal{S}_1\rangle\mid \mathcal{T}_1\rangle\mid
C_{max1}\rangle+\mid \mathcal{S}_2\rangle\mid
\mathcal{T}_2\rangle\mid C_{max2}\rangle)$

\hspace{0.3in}$=1/\sqrt{2}( \mid \underline{00}0001 \rangle\otimes
\mid \underline{01}0001 \rangle\otimes \mid \underline{00}0110
\rangle\otimes \mid \underline{11}0100 \rangle\otimes \mid
\underline{10}0011 \rangle\otimes \mid \underline{01}0111
\rangle\otimes \mid \underline{10}0011 \rangle$

\hspace{1.0in}$\otimes \mid \underline{01}0001 \rangle\otimes\mid
0001000\rangle\otimes\mid 0001000\rangle\otimes\mid
0000110\rangle\otimes\mid 0000100\rangle\otimes\mid 0001000\rangle
$

\hspace{0.3in}$+ \mid \underline{01}0011 \rangle\otimes \mid
\underline{00}0010 \rangle\otimes \mid \underline{10}0101
\rangle\otimes \mid \underline{11}0100 \rangle\otimes \mid
\underline{10}0011 \rangle\otimes \mid \underline{01}0111
\rangle\otimes \mid \underline{00}0101 \rangle$

\hspace{1.0in}$\otimes \mid \underline{01}1010 \rangle\otimes\mid
0000111\rangle\otimes\mid 0010100\rangle\otimes\mid
0001000\rangle\otimes\mid 0000100\rangle\otimes\mid
0010100\rangle)$

 \medskip
 \section{Searching for a Schedule with a Given Makespan}
 \label{GeneralizedGroverSearchAlgorithms}
 \medskip

In our approach, a generalized version of Grover's search
algorithm allows us to find a schedule with a given makespan. The
basic ideas of Grover's quantum search algorithm are discussed
next.

\medskip

Consider a search space $\mathcal{T}_{search} = \{E_{x} \}$
consisting of $N=2^{n}$ elements. Each element $E_{x}, 1 \le x \le
2^{n} $, is uniquely identified by a binary $n$-tuple $x$, called
{\it the index} of the element. We assume that $M \le N$ elements
satisfy the requirements of a query and we wish to identify one of
them.

\begin{figure}[h]
\begin{center}
\includegraphics[width=12cm]{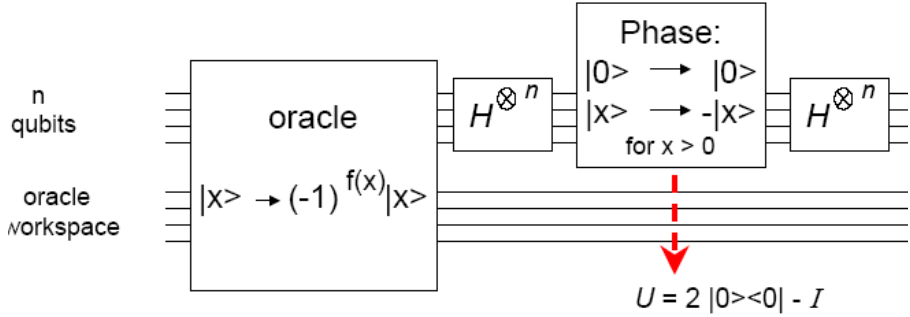}
\end{center}
\caption{A quantum circuit for the Grover's iteration}
\label{Grover2}
\end{figure}

The classic approach is to repeatedly select an element $E_{j}$,
decide if the element is a solution to the query, and if so,
terminate the search. If there is a single solution  ($M=1$) then
a classical exhaustive search algorithm requires
$\mathcal{O}(2^{n})$ iterations.

For the Grover quantum searching algorithm, we apply a
Walsh-Hadamard transform to create an equal superposition state
which includes all elements of the search space:

$$
\mid \psi \rangle = { 1 \over \sqrt{N}} \sum_{x=0}^{N-1} \mid x
\rangle.
$$
Then we perform Grover's iterations. The circuit for this
algorithm is presented in Figure \ref{Grover2}

An oracle examines an index/label, x, and decides if it matches
the search argument or not. To abstract this process we consider a
function $f(x)$ with $0 \le x \le 2^{n}-1$ such that

$$
 f(x) =
 \left\{
 \begin{array}{ll}
 0 & \text{if}~~ x~ \text{is~not~a~solution} \\
 1 & \text{if}~~ x~ \text{is~a~solution}.
 \end{array}
 \right.
$$
An {\it oracle qubit}, $ \mid q \rangle$, initially set to $\mid 0
\rangle$ is reset to $\mid 1 \rangle$ when the oracle recognizes a
solution to the search problem we pose. The black box oracle $O$
performs the following transformation

$$
 O\mid x \rangle \mid q \rangle = \mid x \rangle \mid q
 \oplus f(x) \rangle.
$$
The oracle qubit\index{oracle qubit} can be initially in the state
$\mid q \rangle = (1 / \sqrt{2}) (\mid 0 \rangle - \mid 1
\rangle). $

Thus this transformation can be rewritten as

$$
O\mid x \rangle ~(\mid 0 \rangle - \mid 1 \rangle ) / \sqrt{2}
=
 (-1)^{f(x)} \mid x \rangle ~(\mid 0 \rangle - \mid 1 \rangle ) /
 \sqrt{2}.
$$
The state of the oracle qubit does not change and can be omitted
from the description of the quantum search algorithm

$$
\mid x \rangle  ~~~~\mapsto~~~~ (-1)^{f(x)} \mid x \rangle.
$$
Let $U$ be the following transformation:

$$
U = 2 \mid 0\rangle \langle 0 \mid - I.
$$
Then a conditional phase shift in Figure \ref{Grover2} applied to
the system is:

$$
S_{p} = H^{\otimes n} U H^{\otimes n} = H^{\otimes n} (2 \mid 0
\rangle \langle 0 \mid - I) H^{\otimes n} = 2 \mid \psi \rangle
\langle \psi \mid - I.
$$
A Grover's iteration consists of $O$, the transformation performed
by the oracle followed by a conditional phase shift:

$$
G = S_{p} O = ( 2 \mid \psi \rangle \langle \psi \mid - I) O
$$

Thus, the quantum search algorithm could be written as:

$$
G^R{ 1 \over \sqrt{N}} \sum_{x=0}^{N-1} \mid x \rangle\mid
q\rangle = [(2\mid \psi \rangle \langle \psi \mid - I) O]^R
{1\over \sqrt{N}} \sum_{x=0}^{N-1}\mid x \rangle(\mid 0 \rangle -
\mid 1 \rangle ) / \sqrt{2} \approx \mid x_0\rangle(\mid 0 \rangle
- \mid 1 \rangle ) / \sqrt{2}
$$
When after $R = \mathcal{O} (\sqrt{N \over M})$ iterations, we
measure the first $n$ qubits and obtain $x_0$,  the solution to
the search problem.

\bigskip

{\it Amplitude Amplification} represents a generalization of the
Grover's quantum search idea \cite{Brassard00,Hoyer00a}. Let
$\mathcal{A}$ be a unitary operator in a Hilbert space,
$\mathcal{H}_{N}$ with an orthonormal basis $ \mid 0 \rangle, \mid
1 \rangle, \ldots \mid N-1 \rangle $; the only condition imposed
on $\mathcal{A}$ is  to be invertible, thus $\mathcal{A}$ must not
involve any measurements.

If $\chi: \{ 0, 1, \dots N-1 \} \mapsto \{ 0,1 \} $ is a Boolean
function we say that the basis state $\mid x \rangle $ is a
``Good'' state if $\chi(x) = 1$ and $\mid x \rangle $ is a ``Bad''
state" if $\chi(x) = 0$. The central piece of the amplitude
amplification is an operator $\mathcal{Q}$ defined as:

$$
\mathcal{Q} = \mathcal{Q}( \mathcal{A}, \chi, \phi, \varphi) =
 - \mathcal{A} S_{0}(\phi) \mathcal{A}^{-1} S_{\chi}(\varphi).
$$
with $\phi$ and $\varphi$ two angles such that $0 \le \phi,
\varphi \le \pi$ and $S_{\chi}$ an operator which conditionally
changes the amplitudes of ``Good'' states:

$$
\mid x \rangle \mapsto
 \left\{
 \begin{array} {r c l}
 e^{i \varphi} \mid x \rangle & \text{if} & \chi(x) = 1 \\
 \mid x \rangle & \text{if} & \chi(x) = 0.
 \end{array}
 \right.
$$
Similarly, $S_{0}$ amplifies the amplitude by a factor $e^{i
\phi}$ if the state is not $\mid 0 \rangle$.

Let $a$ denote the probability of finding a ``Good'' element $x$;
amplitude amplification allows to find a ``Good'' $x$ after an
expected number of applications of $\mathcal{A}$ and of the
inverse of $\mathcal{A}$; the number of iterations is proportional
to $1/\sqrt{a}$. We also define the angle $\theta$ such that:

$$
\sin^{2} (\theta) = a.
$$
Grover's algorithm is a particular instance of amplitude
amplification when the oracle implements the Boolean function
$f=\chi$, and the transformation $\mathcal{A}$ is the
Walsh-Hadamard transform $W=H^{\otimes n}$ on n qubits.

This iteration carried out by transformation $Q$ can be regarded
as a \emph{rotation} in the two-dimensional space spanned by the
the state of a uniform superposition of non-solutions and the
state consisting of a uniform superposition of solutions to the
search problem. The initial state may be expressed as:

$$
\mid \psi_0 \rangle=\sqrt{a}\mid Good\rangle+\sqrt{1-a}\mid
Bad\rangle$$

\begin{figure}[h]
\begin{center}
\includegraphics[width=14cm]{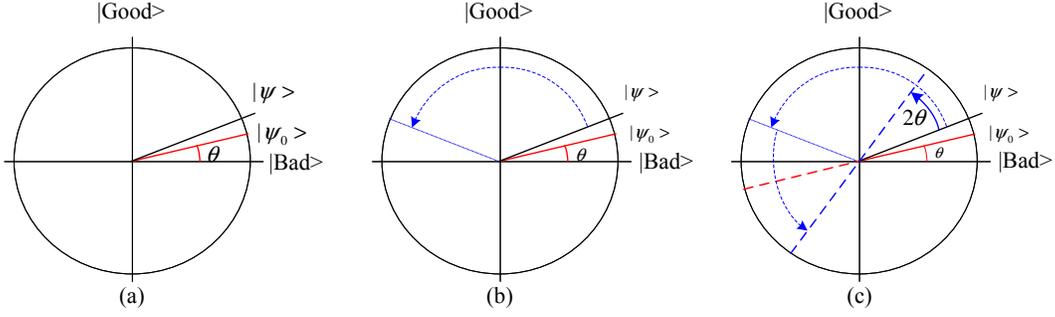}
\end{center}
\caption{The search operator $Q$ performs a rotation toward $\mid
Good\rangle$ states by $2\theta$ radians. $\theta$ is the angle
between the initial state $\mid \psi_0\rangle$ and the $\mid Bad
\rangle$ in the two-dimensional space spanned by the $\mid
Good\rangle$ and $\mid Bad\rangle$ states and $\sin^{2} (\theta) =
a$ with $a$ the probability of finding a ``Good'' element $x$. (a)
The current state $\mid \psi\rangle$ and the initial state $\mid
\psi_0\rangle$. (b) The oracle operation $S_{\chi}$ performs a
reflection of the current state $\mid \psi\rangle$  about the
vector $\mid Good \rangle$. (c) $\mathcal{A}S_0\mathcal{A}^{-1}$
performs a reflection about the initial state $\mid \psi_0\rangle$
} \label{amplitude}
\end{figure}

Figure \ref{amplitude} presents the effect of the transformation
$Q= -\mathcal{A} S_0 \mathcal{A}^{-1}S_{\chi}$ as:

\begin{itemize}

\item the oracle operation $S_{\chi}$ performs a reflection about the
vector $\mid Good \rangle$.

$$
S_{\chi}\mid x\rangle=\mid x\rangle ~~(\chi(x)=1)~~~~~~~~~~~~S_{\chi}\mid x\rangle=-\mid x\rangle ~~ (\chi(x)=0)
$$
\item $\mathcal{A}S_0\mathcal{A}^{-1}$ performs a reflection about the initial state
$\mid \psi_0\rangle$

$$S_0\mid 0\rangle=\mid 0\rangle ~~~~~~~~~~~~~~S_0\mid x\rangle=-\mid x\rangle ~~ (x\neq0)
$$

\item $Q= -\mathcal{A}S_0 \mathcal{A}^{-1}S_{\chi}$ performs a rotation toward $\mid
Good\rangle$ vector by $2\theta$ radians, where $\sin^2 \theta=a$

\end{itemize}

Each iteration $Q$ will rotate the system state by $2\theta$
radians toward the solutions of the searching problem. Thus after
$m$ iterations, the measurement on the final state
$Q^m\mid\psi_0\rangle$ will produce a ``Good'' state with
probability equal to $\sin^2((2m+1)\theta)$. The amplitude
amplification  algorithm could find a ``Good'' solution in
$\mathcal{O}({1\over \sqrt{a}})$ \cite{Brassard00,Hoyer00a}. Each
iteration involves the application of  $\mathcal{A}$ and
$\mathcal{A}^{-1}$.

\medskip

Now let us return to our scheduling problem and recall that:

\begin{itemize}

\item

We prepare each job vector $\mid J_i \rangle$ in a superposition
state which includes the running times on all machines. The first
$m$ qubits of a job vector are used for the index of the machine
and remaining $q$ qubits are used for the running time of that job
on the machine.

\item

We summarize the execution time of all jobs according to their
machine indexes, which produces all schedules and gives the
running time of each machine $T_j$ under these schedules. The
system is prepared in an entangled state:

$$\mid
\overbrace{\mathcal{S}\mathcal{T}}\rangle = {1 \over
\sqrt{\sigma}}\sum_{\text{each schedules
k}}(\bigotimes_{i=1}^{N}\mid J_{ik}\rangle\bigotimes_{j=1}^{M}\mid
T_{jk}\rangle ) ~~~~~~~~\sigma= 2^{m 2^{n}},
$$
a superposition of job vectors and running time vectors of all
possible schedules.

\item

We obtain the maximum running time among all machines using the
{\tt Max} quantum circuit and prepare the system in state:

$$
\mid \overbrace{\mathcal{S}\mathcal{T}C_{max}} \rangle  = {1 \over
\sqrt{\sigma}}\sum_{\text{each schedules
k}}(\bigotimes_{i=1}^{N}\mid J_{ik}\rangle\bigotimes_{j=1}^{M}\mid
T_{jk}\rangle \mid C_{max~k}\rangle) ~~~~~~~~\sigma= 2^{m 2^{n}}$$

\end{itemize}
\bigskip

As we can see, our algorithm successfully prepares the system in
an equal superposition of all $2^{m 2^{n}}$ possible schedules. We
define this whole preparation process as $\mathcal{Z}$. This
$\mathcal{Z}$ transformation does not carry out a measurement of
the system at any time. Therefore, there exists an inverse
transformation operation $\mathcal{Z}^{-1}$. We can use the
amplitude amplification algorithm to search the schedule with the
makespan $D_{k}= \mu$. If we find such a makespan, the job vectors
will be projected as well. These projections will give us the
actual mapping of jobs to machines corresponding to the schedule
with the given makespan.

The searching process consists of the following steps:

\begin{itemize}

\item Apply the $\mathcal{Z}$ transformation on $\mid 0\rangle$ to
prepare the system in an equal superposition of all $2^{m 2^{n}}$
possible schedules, $\mid \psi\rangle$.

\item

Repeat the following steps $\mathcal{O}(\sqrt{\sigma})$ times:

\hspace{.5in}Apply $Q$ on $\mid \psi\rangle$, in which $Q =
-\mathcal{Z}S_0 \mathcal{Z}^{-1}S_{\chi}$

\item

Measure the resulting state.

\item

Return the result; the job vectors give the detailed schedule.
\end{itemize}
\medskip

The oracle in our searching algorithm exhibits some difference
with the oracle in Grover's algorithm which checks all qubits to
reverse the solution(s) of the searching problem. In our case, the
oracle only checks a subset,  $C_{max}$ qubits. It is easy to
implement such an oracle using an idea similar to the one in
\cite{Nielsen2000}.

The algorithm presented in this paper can be optimized; to avoid
over rotation we could use the fix-point quantum search
\cite{Grover05}.

\medskip
\section{Scheduling Problems with a Quantum Counterpart}
\medskip

Many other scheduling problems can be reformulated to take
advantage of quantum search and exhibit a square root speedup
versus their classical counterparts. Such problems require that a
synthetic measure of performance, $\mu$ be within a given range,
$\mu_{min} \le \mu \le \mu_{max}$.

The generic quantum algorithm proceeds as follows:

\begin{enumerate}

\item
Devise an encoding scheme for the information required to compute
$\mu_{S_{i}}$ for a given schedule $S_{i}$.

\item
Design an algorithm to compute the synthetic measure of
performance, $\mu_{S_{i}}$.

\item
Construct $\mid \mathcal{Q} \rangle$ with
$\mathcal{Q}=(\mu_{S_{i}},S_{i})$ in a superposition state for all
possible schedules.

\item
Design the quantum circuits for the specific encoding scheme and
for the algorithms. Given a set of values $\{q_{1}, q_{2}, \ldots
q_{n}\}$ the circuit should be able  to compute a function
$\mu_{S_{i}}=f(q_{1}, q_{2}, \ldots q_{n})$. For example, we may
wish to compute the maximum, the minimum, or an average value.

\item
Design an oracle to identify a specific value of the function
$\mu_{S_{i}}$.

\item Use the quantum search to find if there is a value
$\mu_{S_{i}}= \mu_{min}$. If so, determine the corresponding
schedule $S_{i}$. Continue this process until $\mu_{S_{i}}=
\mu_{max}$.

\item
If no schedule can be found then report failure, otherwise provide
the list of all schedules $S_{i}$ and the corresponding measures
of performance, $\mu_{S_{i}}$.

\end{enumerate}

Consider for example the $R||\sum_{i=1}^{N}C_i^S$ scheduling
problem when the goal is to optimize the average completion time
in unrelated parallel machine environment. It has the similar
encoding process as the $R||C_{max}$ problem. During the process
we construct all schedules, we also summarize the completion time
of different jobs using some simple arithmetic circuits, followed
by Grover-type search. Other scheduling problems such as
minimizing the average waiting time, could also take advantage of
quantum search.

Oftentimes, we have to search for schedules that optimize the
largest subset of a set of synthetic measures of performance,
$\mu,\nu, \pi, \rho, \theta \ldots$. For example, we could have
multiple synthetic performance indicators related to: timing,
resource utilization, cost, and quality of service. In this case
we would run repeatedly the scheduling algorithm for each
performance measure and search for a solution in the given range
for each measure. Once we have conducted all individual searches
we determine the intersection of all schedules that satisfy all
conditions; if the set is empty we remove individual conditions
one by one until we find the a non-empty set.

Scheduling is also intimately related to planning when we have a
complex goal and the sequence of actions to reach each goal have
to be determined. The scenario described above is repeated for
each plan thus the square root speedup of algorithms based upon
quantum search becomes even more important.

\medskip
\section{Summary}
\medskip

When a deadline is imposed, or when we wish to find a schedule
with a given range of possible average completion time we discover
that  a full range of scheduling problems have a quantum
counterpart which can take advantage of Grover's search.

Many scheduling problems, resource allocations, and path-finding
problems, share some common properties with the $R||C_{max}$
problem discussed in this paper: a well-defined initial state, a
well-defined desired state or a range of desired states, many
paths to reach the desired state, and well-defined attributes of
an optimal path.

The quantum algorithmic solution to such problems requires the
following steps:

\begin{itemize}

\item Prepare the initial state in an equal superposition of all
possible choices.

\item Use some reversible quantum arithmetic to compute the
specialized property (makespan in our case) needed.

\item Construct the necessary oracle circuit.

\item Use Grover-type algorithms to search for the desired solution.

\end{itemize}

The solution we propose based upon Grover's algorithm is not
universally applicable. Problem requiring specific optimization
criteria may require quantum circuits that cannot be easily
implemented with reversible quantum arithmetic. Moreover,
Grover-type algorithms lead to a square-root improvement over the
exhausting search, while many good classical algorithms may have
better performance for some special problems.

\section{Acknowledgments}

The research reported in this paper was partially supported by
National Science Foundation grant CCF 0523603. The authors express
their thanks to Lov Grover, Pawel Wocjan,  and an anonymous
reviewer for their comments and suggestions.

\end{document}